# Excitonic properties of semiconducting monolayer and bilayer $MoTe_2$

**C. Robert[1a], R. Picard[1], D. Lagarde[1], G. Wang[1], J. P. Echeverry[1], F. Cadiz[1], P. Renucci[1], A. Högele[2], T. Amand[1], X. Marie[1], I. C. Gerber[1b] and B. Urbaszek[1]**

[1] *Université de Toulouse, INSA-CNRS-UPS, LPCNO, 135 Ave. de Rangueil, 31077 Toulouse, France*
[2] *Fakultät für Physik and Center for NanoScience (CeNS), Ludwig-Maximilians-Universität München, Geschwister-Scholl-Platz 1, 80539 München, Germany, EU*

**Abstract.**

$MoTe_2$ belongs to the semiconducting transition metal dichalcogenide family with some properties differing from the other well-studied members $(Mo,W)(S,Se)_2$. The optical band gap is in the near infrared region and both monolayers and bilayers may have a direct optical band gap. We first simulate the band structure of both monolayer and bilayer $MoTe_2$ with DFT-*GW* calculations. We find a direct (indirect) electronic band gap for the monolayer (bilayer). By solving the Bethe-Salpeter equation, we calculate similar energies for the direct excitonic states in monolayer and bilayer. We then study the optical properties by means of photoluminescence (PL) excitation, time-resolved PL and power dependent PL spectroscopy. We identify the same energy for the B exciton state in monolayer and bilayer. Following circularly polarized excitation, we do not find any exciton polarization for a large range of excitation energies. At low temperature (T=10 K), we measure similar PL decay times of the order of 4 ps for both monolayer and bilayer excitons with a slightly longer one for the bilayer. Finally, we observe a reduction of the exciton-exciton annihilation contribution to the non-radiative recombination in bilayer.

[a] cedric.robert@insa-toulouse.fr [b] igerber@insa-toulouse.fr

## I. Introduction

Monolayers (ML) of group VI transition metal dichalchogenide (TMD) form a new class of semiconducting materials with exciting properties for electronics and optoelectronics applications[1–4]. $MoS_2$ was the first studied material[5,6] quickly followed by extensive studies on binary $WS_2$, $MoSe_2$, $WSe_2$[7,8] and ternary alloys[9–11]. In the 2H hexagonal structure, these materials all share common properties: (i) an indirect to direct band gap crossover when the material is thinned down to the monolayer limit[5,6], where the direct gap is located at the $K^{+/-}$ points of the hexagonal Brillouin zone, (ii) strong binding energies for excitons (Coulomb bound electron-hole pairs) of several hundreds of meV [12–14], and (iii) spin and valley dependent optical selection rules due to the lack of crystal inversion symmetry and large spin-orbit coupling[15–19]. Nevertheless, several differences in the physical properties between semiconducting TMD ML materials can be pointed out. The amplitude of the valence spin-orbit splitting varies from less than 200 meV for $Mo(S,Se)_2$ ML to more than 400 meV for $W(S,Se)_2$ ML. The sign of the splitting between intravalley bright and dark excitons is the opposite in $MoSe_2$ and $WSe_2$ ML[20], dramatically affecting their optical properties[11,21,22]. Finally, despite comparable structural and optical quality, the valley/spin polarization properties probed in optical spectroscopy are very different between materials. For $MoS_2$ ML, the degree of circular polarization of the photoluminescence (PL) decreases monotonously when the detuning of the excitation laser energy relative to the excitonic ground state increases[23,24]. For $WSe_2$ and $WS_2$ ML, circular polarized PL has been reported even for excitation energies far from the resonance but the polarization degree is exalted for excitation energies in resonance with exciton excited states (2s or 2p)[14,25]. In contrast, no PL polarization has been observed in $MoSe_2$ ML except for quasi-resonantly excitation[24].

Recently, the family of semiconducting TMD monolayers expanded with a fifth binary material: $MoTe_2$[26,27]. Two striking properties have been highlighted by the first studies on this material: the optical band gap of $MoTe_2$ ML is in the near infra-red (1.1 eV at room temperature[26]) whereas it lies in the red part of the visible spectrum for S and Se-based compounds. Secondly, the luminescence yield of $MoTe_2$ bilayers (BL) is in the same order of magnitude as the luminescence yield of the ML opening a debate on a possible direct band gap for BL[27,28]. Nevertheless, the high luminescence yield of BL is up to now the only argument pointing towards a direct band gap. In addition, many properties such as radiative lifetime, spin/valley polarization, energy of exciton excited states have not been measured in this material yet. In this paper we partially fill this gap by studying theoretically and experimentally the optical properties of both $MoTe_2$ ML and BL. The paper is organized as follows. In the next section, we use DFT+*GW* calculations to simulate the band structures of both ML and BL in a single particle picture (no excitonic effects). Remarkably we find an indirect electronic band gap for the BL. We then solve the Bethe-Salpeter equation (BSE) to include the strong electron-hole Coulomb attraction when calculating the energy and the oscillator strength of direct excitonic transitions. Section III is dedicated to stationary PL measurements. Experimentally, we measure the splitting between A and B excitons by PL excitation spectroscopy and compare the results with $G_0W_0$+BSE calculations. We also do not detect any measurable PL polarization for excitation energy as close as 60 meV above the energy of the A-exciton suggesting a behaviour similar to $MoSe_2$ ML. In section IV, we use time-resolved PL to measure the exciton lifetime in the range of a few ps and find a slightly longer one for BL. Finally, in section V, we show that exciton-exciton annihilation is larger in ML than in BL.

## II. Electronic Band Structure Calculation

In recent years, DFT+*GW* methods have been successfully applied to calculate the electronic band gap of 2D semiconductors materials ($E_G$), see for instance [14,29–32]. In the present study, the exploration of the electronic structure and optical properties of ML and BL $MoTe_2$ has been performed using the VASP code [33,34]. It uses the plane-augmented wave scheme[35,36] to treat core electrons, when fourteen electrons for Mo and six for Te ones are explicitly included in the valence states, with a plane-wave energy cut-off of 400 eV. The Heyd-Scuseria-Ernzerhof (HSE) hybrid functional[37–39] is used to build the needed wavefunctions, based on 600 electronic states, to calculate the full-frequency-dependent quasiparticle band structure at the $G_0W_0$ level of theory[40], including spin-orbit coupling (SOC) but not the excitonic effects at this stage. This particular choice of the computational settings has been discussed in detail in Ref [20]. A grid of (12 × 12 × 1) *k*-points has been used, in conjunction with a vacuum height of at least 17 Å for both the ML and BL systems. For the latter, we have selected the stacking geometry of AA' type, since it appears to be the most stable BL structure[41]. It corresponds to the point group $D_{3d}$ symmetry, with and eclipsed stacking with Mo over Te. The optimized interlayer distance is 7.00 Å, when Van der Waals forces are taken into account, via the optB86b-VdW scheme[42].

In Fig.1a, the resulting DFT+$G_0W_0$ band structure of the $MoTe_2$ ML is shown after a Wannier interpolation procedure performed by the WANNIER90 program[43]. Its main features agree well with previous theoretical studies [29,44]. It has a direct electronic band gap in K-valley with a value of 1.72 eV, a SOC splitting in the valence band of 275 meV and it is -58 meV in the conduction band. The negative sign for the conduction band SOC means that both conduction band minimum and valence band maximum have the same spin. In comparison the BL band structure at the same level of theory is given in Fig.1b. If the direct $K_v$-$K_c$ gap remains almost the same with a value of 1.66 eV for the BL, now it appears that the indirect $K_v$-$T_c$ quasiparticle band gap is the lowest one with a value of 1.60 eV. Thus the interlayer interaction leads to a transition from a direct to an indirect band gap similar to other group VI TMD[41] with an energy separation between $K_c$ and $T_c$ remaining small compared to others

TMD systems. Nevertheless, contrarily to $MoS_2$ BL, the valence band minimum remains in the K valley[41]. The interlayer interaction also tends to enhance the energy separation between the two highest valence states of different spin for each layer, whereas the opposite is true for the lowest unoccupied ones. As a result, the SOC splittings become 304 and -46 meV for the valence and conduction bands, respectively.

With DFT+*GW* calculations we find that the electronic band gap is direct for the ML and indirect for the BL. But in TMD materials, the optical properties are not governed by the band to band recombinations. Indeed, due to the very large exciton binding energy ($E_B$~500 meV), the PL spectrum is dominated by the ground exciton transition (also called optical band gap $E_{PL}=E_G-E_B$). To calculate the optical band gap, we need to include the electron-hole Coulomb interaction into the model, by solving the BSE. Practically BSE spectra, are obtained in the Tamm-Dancoff approximation, by using the six highest valence bands and the eight lowest conduction bands for $MoTe_2$ ML and 12 valence and 16 conduction bands for the BL to obtain eigenvalues and oscillator strengths, with a complex shift of 25 meV to broaden the theoretical absorption spectra. Only direct transitions are taken into account in this calculation. For TMD ML, this is assumed to give a good description of the main excitonic states as both electronic and optical band gaps are direct. But for $MoTe_2$ BL, for which we find an indirect electronic band gap, we would need to add the exciton-phonon interaction into the scheme to conclude on the indirect-direct nature of the optical band gap. Unfortunately this refinement is not trivial and is far beyond the scope of this study[45].

The BSE spectra are given in Fig.1c and Fig.1d. We can identify in both spectra the ground states of both A and B excitons (namely $X_A^{1s}$ and $X_B^{1s}$). Interestingly the position of the $X_A^{1s}$ and $X_B^{1s}$ peaks remains at the same energy when going from ML to BL. The corresponding binding energy is thus 0.46 eV for the A-exciton in ML and 0.4 eV in BL. From the ML spectrum we can also identify a small shoulder at 0.04 eV below the $X_B^{1s}$ peak, and can be safely assigned to a transition associated to the *2s* state of the A exciton. This peak is also present in the BL spectrum, but it is slightly shifted, by 0.01 eV to lower energy. Another interesting feature is the extra peak located at 0.01 eV above the $X_A^{1s}$ peak when stacking the two layers. This transition has non-negligible oscillator strength and is of interlayer character. Indeed it involves the 5$^{th}$ and 6$^{th}$ conduction bands which clearly possess a delocalized character over the two layers and is composed mainly of $d_{xz}$ and $d_{yz}$ orbitals of the two Mo atoms. Note that this kind of weak interlayer transitions in the vicinity of the A peak, have been previously reported in the case of $MoS_2/WS_2$ hetero-structure[30] but also for bilayers[41].

## III. Continuous-wave spectroscopy

Our DFT calculations predict at the *GW* level a direct gap for the ML and an indirect gap for the BL with a direct gap only 60 meV above in energy. This energy difference is very small once excitonic effects (~500 meV) are taken into account. This motivates the optical spectroscopy studies described in this section, where ML and BL emission energies and intensities can be determined.

For experimental studies of optical properties, we use $MoTe_2$ flakes obtained by micro-mechanical cleavage of a bulk $MoTe_2$ crystal (supplied by the company 2D semiconductors) on 90 nm $SiO_2$/Si substrate using viscoelastic stamping[46]. The ML and BL regions are identified by optical contrast (see inset of Fig.2a) and very clearly in PL spectroscopy (Fig.2a). Experiments between 4 K and 300 K are carried out in a confocal microscope optimized for polarized PL experiments in the near infra-red (IR). After dispersion with a near IR blazed grating, the PL is analysed with an InGaAs photodiode array for continuous-wave (cw) experiments or a S1 photocathode streak camera (Hamamatsu C5680) for time-resolved photoluminescence (TRPL) measurements. For cw excitation, three different lasers are used. A standard HeNe laser is used for strongly non-resonant excitation whereas a cw Ti:Sa laser and a tunable laser diode are used to adjust the excitation wavelength from 750 nm to 1000 nm. For TRPL measurements, we excite the sample with 1.5 ps pulses of a Ti:Sa laser at a wavelength of 850 nm and a repetition rate of 80 MHz. For all measurements, the laser spot diameter is around 1 µm on the sample (i.e. much smaller than the flake size) and the average power is set below 100 µW.

Fig.2a presents the low temperature PL spectra of both ML and BL for an excitation energy above the free carrier gap that we calculate by DFT-*GW* (Fig.1a-b). The spectra are composed of two main peaks typically attributed to the ground states of the neutral exciton $X_A^{1s}$ (at 1.184 eV for the ML) and trion T (at 1.159 eV for ML) for the A transition. The energies of the peaks are in good agreement with the previously reported PL spectra of $MoTe_2$ ML and BL[26–28]. For the $MoTe_2$ ML, we measure a separation between $X_A^{1s}$ and T peaks of 25 meV in agreement with the measurement of the binding energies of positively and negatively charged excitons in a field effect structure[44] (24 meV and 27 meV respectively). For the BL, the separation between peaks is smaller (18 meV) but we cannot unambiguously attribute the low energy peak to the trion signature as no charge tunable device based on BL has been reported yet. The full width at half maximum (FWHM) is 7 meV for both $X_A^{1s}$ and T in ML indicating an optical quality as high as for the best $MSe_2$ (M=Mo, W) ML[47]. For comparison, the smallest linewidth at low temperature we obtained on chemically treated $MoS_2$ ML is typically 15 meV[48]. Thus, $MoTe_2$ material is particularly suitable to study the complex exciton/trion fine structures of TMD ML. In addition to the two main peaks we observe features on the low energy part of the spectrum and between $X_A^{1s}$ and T peaks. We attribute them to complex localized states (marked as Loc on Fig.2a) as their contribution to the PL spectrum vanishes when the temperature increases (see Fig.2b). We also want to point out that the amplitude of these localized states varies from ML to ML making the origin of these features difficult to attribute at this stage.

Several important characteristics of TMD ML (optical generation of valley polarization, second-harmonic generation efficiency) are known to strongly depend on the laser excitation energy as the light matter interaction is strongly enhanced at the excitonic resonances[14,24,49,50]. Therefore, probing the excited excitonic states is of particular importance. Fig.3a presents the PLE spectra corresponding to the variation of the $X_A^{1s}$ PL intensity as a function of the laser energy. A clear resonance of the $X_A^{1s}$ PL peak is observed at 1.45 eV for both ML and BL which corresponds to the signature of the ground B exciton state $X_B^{1s}$. We thus find a splitting B-A of 270 meV for the ML in agreement with the room temperature reflectivity measurement of Ruppert *et al.* (260 meV)[26]. This is also in good agreement with the BSE spectrum of Fig.1c where we calculate a splitting of 330 meV. We notice that this splitting is larger than for the other Mo-based TMD. Finding exactly the same $X_B^{1s}$ resonance energy for both ML and BL is also interesting and in good agreement with the calculations of Fig.1c-d. According to the BSE spectra, we could also expect the same energy for the $X_A^{1s}$ exciton in both ML and BL. Nevertheless, we observe a red shift of 36 meV between the $X_A^{1s}$ PL peak of ML and the exciton PL peak of BL (see Fig.2a). A possible explanation is that the PL peak observed at 1.148 eV in the PL spectrum of BL $MoTe_2$ may not correspond to the $X_A^{1s}$ direct exciton state. In other words, the optical band gap (including excitonic effects) of $MoTe_2$ BL could actually be indirect. The calculation of indirect exciton states would be interesting in the future to possibly validate this hypothesis. Using the ratio of integrated BL PL versus ML PL is not sufficient to distinguish between direct and indirect optical transitions. Indeed, the laser energy $E_{laser}$ used for excitation plays an important role, as can be seen in Fig.3b. For example, the BL PL emission for $E_{laser}$=1.319 eV is more intense than the ML PL, probably because optical absorption in the BL is more efficient at this energy.

Contrary to $WSe_2$ ML[51], we do not find a clear signature of the $X_A^{2s}$ excited state in the PLE spectrum of $MoTe_2$ ML. A possible explanation is that this state is very close to the $X_B^{1s}$ state like in $MoSe_2$ ML[52] and cannot be distinguished in the PLE spectrum. This is supported by the *GW*-BSE calculations of Fig.1c where we find a separation $X_B^{1s}$-$X_A^{2s}$ of 40 meV thus smaller than the width of the PLE peak in Fig.3a. Two-photon experiments would be an efficient way to probe the $X_A^{2p}$ state without being sensitive to the $X_B^{1s}$ state[52]. However this requires exciting the ML with energies lower than 0.75 eV which is beyond the tuning range of conventional laser systems.

We can also point out two interesting differences between PLE spectra of ML and BL. First, the resonance peak at 1.45 eV is clearly broader in the BL case and could be a sign of the $X_A^{2s}$ contribution. Secondly, for excitation energies lower than 1.4 eV, the PL intensity of the neutral exciton peak in BL is larger than for the ML while it is smaller for the excitation energy used in Fig.2a (1.96 eV). This could be an experimental evidence of the interlayer exciton state reported in Fig.1d at 1.38 eV.

TMD ML obey chiral optical interband selection rules[15] that allow for optical excitation in either the $K^+$ or $K^-$ valley depending on excitation laser helicity. For the range of excitation energies shown in Fig.3a, we do not measure any significant circular (linear) polarization following circularly (linearly) polarized excitation. This behavior is in agreement with the very recent zero polarization reflectivity measured at zero magnetic field[53]. This is also similar to what was observed for $MoSe_2$ ML[23]. Only one recent study reports a circular PL polarization degree of 20% in $MoSe_2$ ML for a difference between the laser energy and the energy of the $X_A^{1s}$ emission of $\Delta E = E_{laser} - E[X_A^{1s}]$=60 meV[24]. LA phonon-assisted intervalley scattering was proposed as the cause of depolarization in $MoSe_2$ and $MoS_2$ ML[24]. The larger value of the LA phonon for $MoS_2$ (30 meV)[54] as compared to $MoSe_2$ (19 meV)[55] would explain why polarization can be observed in $MoS_2$ for larger ΔE. Interestingly, the LA phonon energy of $MoTe_2$ is even smaller (12 meV)[56] than that of $MoSe_2$. Thus, only excitation very close to the resonance may initiate valley polarization in this material. Unfortunately, for ΔE<60 meV, a strong Raman scattering signal is superimposed on the PL signal resulting in unreliable polarization measurements.

## IV. Time-resolved measurements

The exciton binding energy in $MoTe_2$ ML can be roughly estimated by taking the difference between the electronic band gap provided by the DFT-*GW* calculation (1.72 eV, see Fig.1a) and the optical band gap measured in PL (1.18 eV, see Fig.2a). We find a binding energy of ~540 meV in agreement with previous estimation of Yang *et al.*[44]. Such a strong exciton binding energy and the associated strong oscillator strength are very interesting properties for strong light-matter coupling studies in ML TMDs[57,58]. Consequently, a short PL emission time can be expected and measuring the exciton radiative lifetime is thus crucial. TRPL is the ideal spectroscopy tool for such a measurement. In our previous works, we measured low temperature PL decay times of 4 ps for $MoS_2$ ML[50] and 2 ps for $MoSe_2$ and $WSe_2$ ML[59]. Our temperature dependent study on $MoSe_2$ ML suggested that this decay time corresponds to the radiative lifetime of excitons before thermalization occurs. In Fig.4, we present the low temperature PL dynamics of $X_A^{1s}$ for both $MoTe_2$ ML and BL. The excitation wavelength of 850 nm (1.458 eV) is chosen to match with the enhanced absorption at the $X_B^{1s}$ exciton transition (see Fig.3a). For the ML, we observe a mono-exponential decay that can be fitted with a characteristic time of $\tau^{ML}$ ~ 3.4 ± 0.5 ps. This dynamic is clearly longer than the time-resolution of our setup measured by detecting the laser pulse backscattered from the sample on the streak camera (shaded area). Following Ref [59], we interpret this fast decay time as the radiative lifetime of excitons. A key argument is that the $X_A^{1s}$ PL intensity is constant in the range 10 K – 40 K (see Fig.2b) excluding any role of non-radiative channels at low temperature. For the BL, the decay is bi-exponential. We attribute the longest decay time (23 ps) to the presence of localized states at the same energy than the free exciton peak. The shortest decay time $\tau^{BL}$ ~ 4.3 ± 0.5 ps can be attributed to the lifetime of free excitons in BL. We want to point out that contrary to the ML, the PL intensity of BL decreases as soon as the temperature is raised. We thus cannot exclude that this decay time is also governed by non-radiative recombination. But interestingly, this time is slightly longer than $\tau^{ML}$ which might hint at the fact that the radiative lifetime of excitons in BL is longer than the radiative lifetime of excitons in ML. Several hypotheses can be proposed at this stage. First it was reported that the optical band gap of $MoTe_2$ BL may be borderline direct/indirect [27,28]. It is thus not surprising to observe a smaller rate of radiative recombination in the BL. Nevertheless, we notice that the decay time we measure for $MoTe_2$ BL is 6 times faster than for the indirect transition in $WSe_2$ BL[60] suggesting a more direct transition in $MoTe_2$ BL. Secondly, potential fluctuations with a high spatial frequency have been proposed to explain the significant discrepancy between the measured exciton lifetime (a few ps) and the theoretical radiative lifetime of free excitons (a few hundreds of fs)[59]. We expect larger Bohr radius excitons in BL to be more sensitive to these fluctuations than smaller Bohr radius excitons in ML and consequently to yield longer radiative lifetime.

## V. Exciton-exciton annihilation

Non-radiative recombination channels are known to play a major role in the poor luminescence yield measured at room temperature in TMD ML[61]. In addition to defect-related recombination, exciton-exciton annihilation (EEA) is known to be very efficient in $MoS_2$, $MoSe_2$, $WS_2$, $WSe_2$ at room temperature even for moderate excitation power density[61,62]. In Fig.5, we plot the variations of the $X_A^{1s}$ PL intensity as a function of excitation power for both ML and BL at 10 K and 200 K. For the ML, the $X_A^{1s}$ PL intensity scales linearly at 10 K whereas it scales sublinearly (~$P^{0.8}$) at 200 K for excitation powers larger than 10 µW. This might be a consequence of the thermal activation of EEA processes. At 10 K, the exciton diffusion is too small and the radiative lifetime is so short that EEA does not compete with radiative recombination. In contrast, when the temperature increases, the mobility of excitons increases[62,63] and so does the effective radiative lifetime of excitons due to thermalization[22,64,65]. This enhances the sensitivity to many-body interactions. Remarkably, for the BL, the situation is different. As shown in Fig.5, the $X_A^{1s}$ PL intensity increases linearly with the excitation power for both temperatures suggesting a reduced EEA rate as compared to ML. We also observe the linearity at room temperature (not shown here). Such a property combined with an efficient optical transition would make $MoTe_2$ BL a very promising candidate for optoelectronics applications requiring high carrier densities including laser or concentrating solar cells[66]. Unfortunately, a clear conclusion cannot be drawn at this stage. First, we notice that at 10 K, the PL intensities of ML and BL are in the same order of magnitude whereas at 200 K, the BL intensity is one order of magnitude lower than the ML intensity (for the same excitation power). This could be due to a higher defect density in BL. Passivation capping or chemical treatments[61,67,68] would help to study the real influence of defects on the optical properties of $MoTe_2$ ML and BL. A second explanation would be that the indirect transition plays a more important role at elevated temperatures (due to thermal activation or direct to indirect crossover). Actually, Yuan and Huang already measured the EEA rates in $WS_2$ ML and BL[69] and found a rate for the BL two orders of magnitude lower than in ML due to the reduced phonon-assisted EEA of indirect excitons in indirect band gap $WS_2$ BL. Thus, we can expect a similar behaviour if $MoTe_2$ BL is indirect. Finally, we cannot exclude that the mobility of excitons is reduced in $MoTe_2$ BL which could explain why the EEA processes are not visible even at high temperatures.

In conclusion, we studied the optical properties of $MoTe_2$ ML and BL. We performed DFT-*GW* calculations and found a direct electronic band gap for the ML and an indirect one for the BL. With one-photon PLE, we found that the energy of the B exciton state is the same in ML and BL in agreement with the BSE calculations. We did not find a clear signature of the $X_A^{2s}$ exciton excited state which may lie close to the $X_B^{1s}$ resonance. We did not detect any circular or linear PL polarization for laser energy as close as 60 meV above the energy of the $X_A^{1s}$ exciton. We measured the exciton lifetimes of ML and BL at low temperature. The lifetime in BL is slightly longer than the radiative lifetime in ML but remains significantly faster than for BL of other TMD materials. Finally, we discussed the observed reduction of the EEA contribution to the non-radiative recombination in $MoTe_2$ BL.

*Acknowledgments*: We thank the ANR MoS2ValleyControl and Programme Investissements d'Avenir ANR-11-IDEX-0002- 02, reference ANR-10-LABX-0037-NEXT and ERC Grant No. 306719 for financial support. A. H. acknowledges the ERC Grant No. 336749. I. C. G. also acknowledges the CALMIP initiative for the generous allocation of computational times, through the project p0812, as well as the GENCI-CINES, GENCI-IDRIS and GENCI-CCRT for the grant x2015096649. I. C. G thanks the CNRS for his financial support. X.M. also acknowledges the Institut Universitaire de France.

**FIG.1a:**

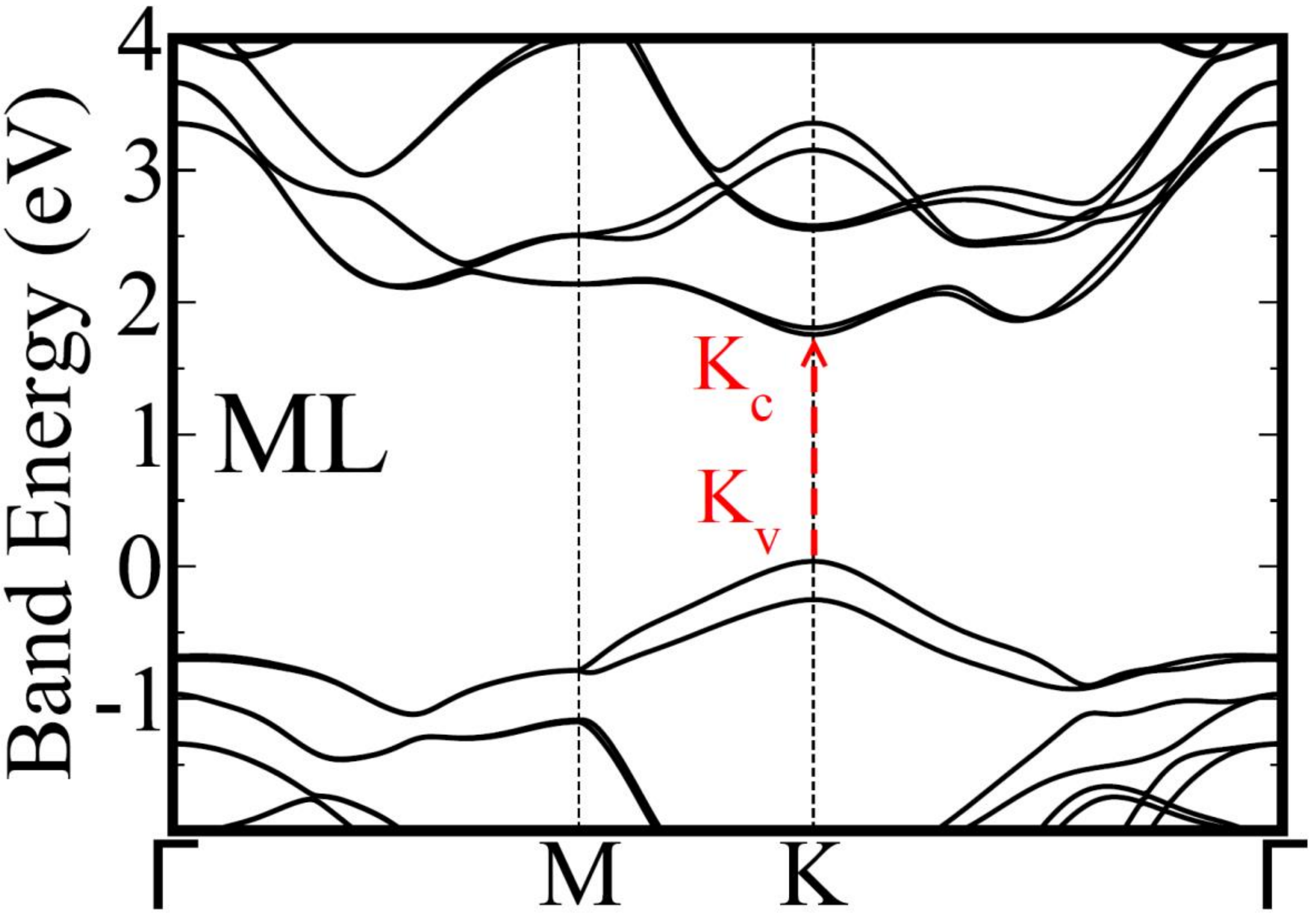

**FIG.1b:**

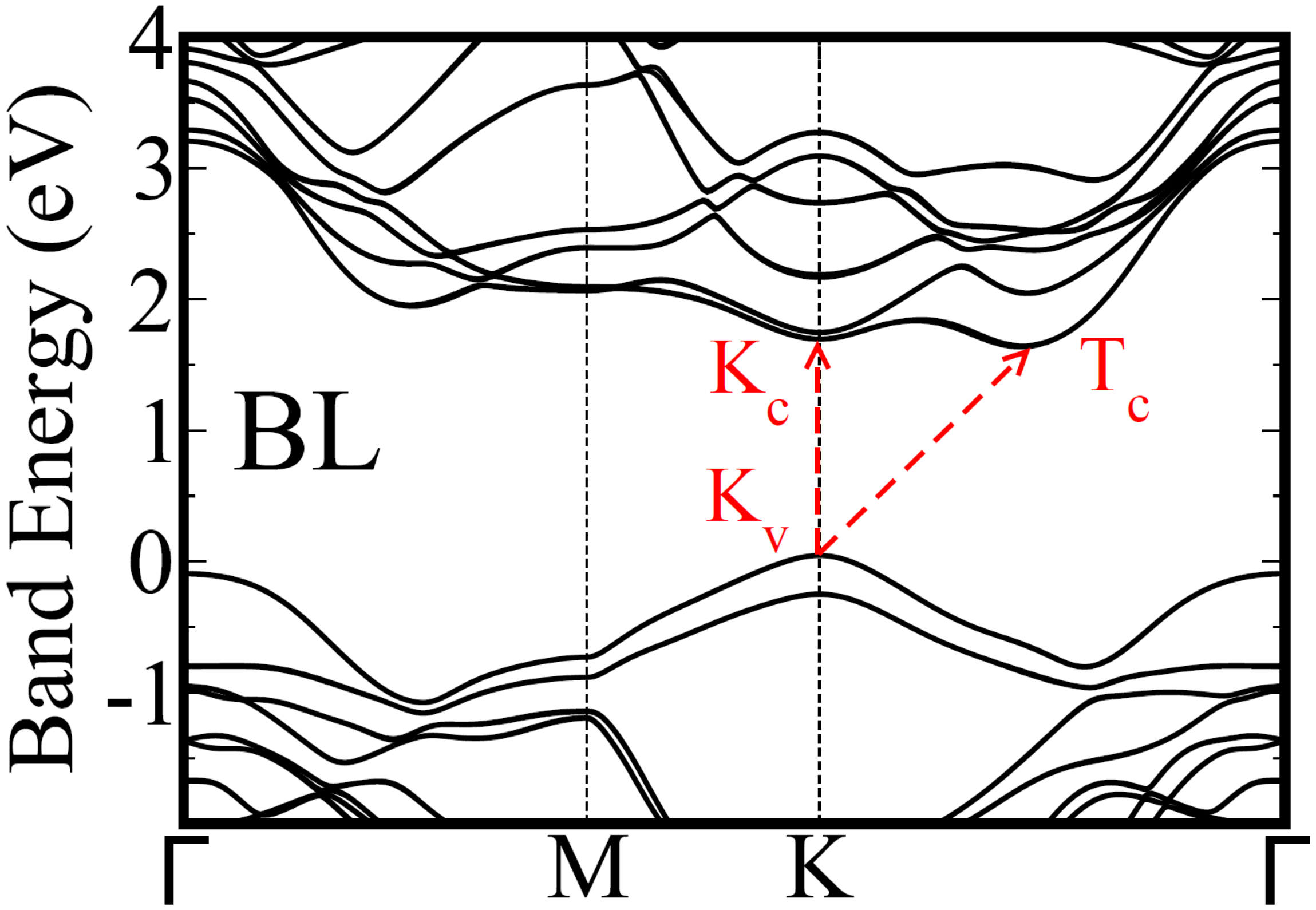

**FIG.1c-d:**

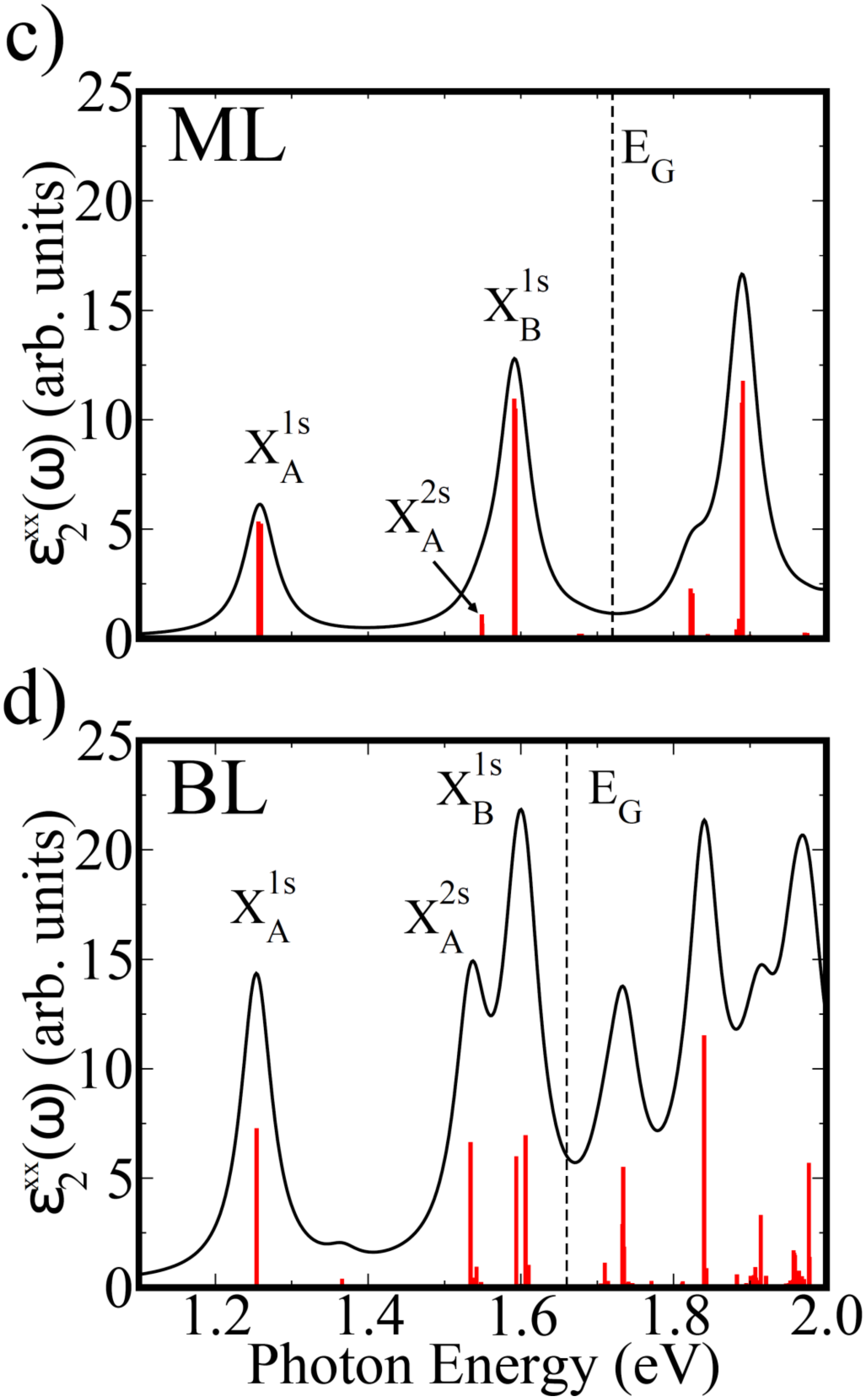

**FIG.2a:**

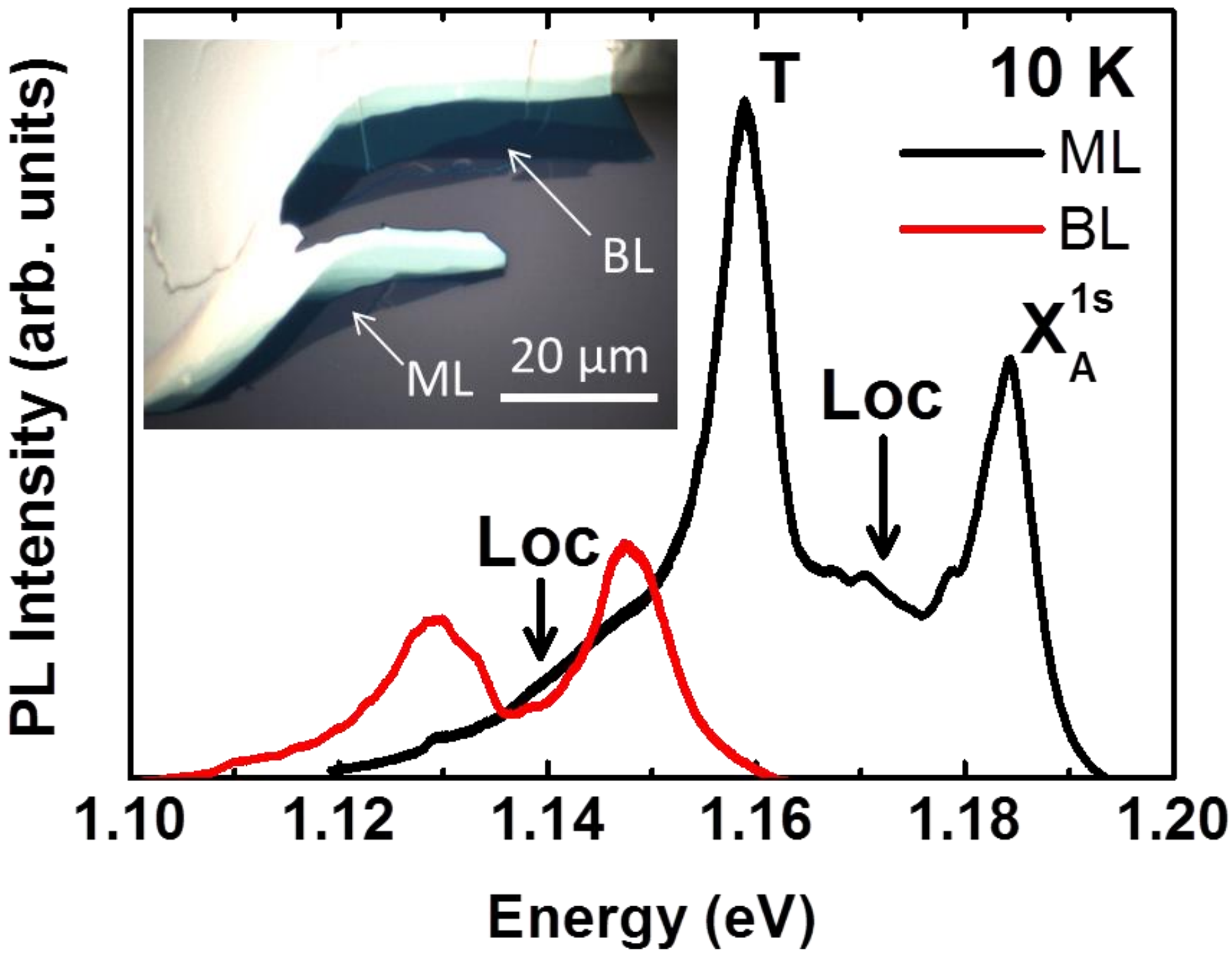

**FIG.2b:**

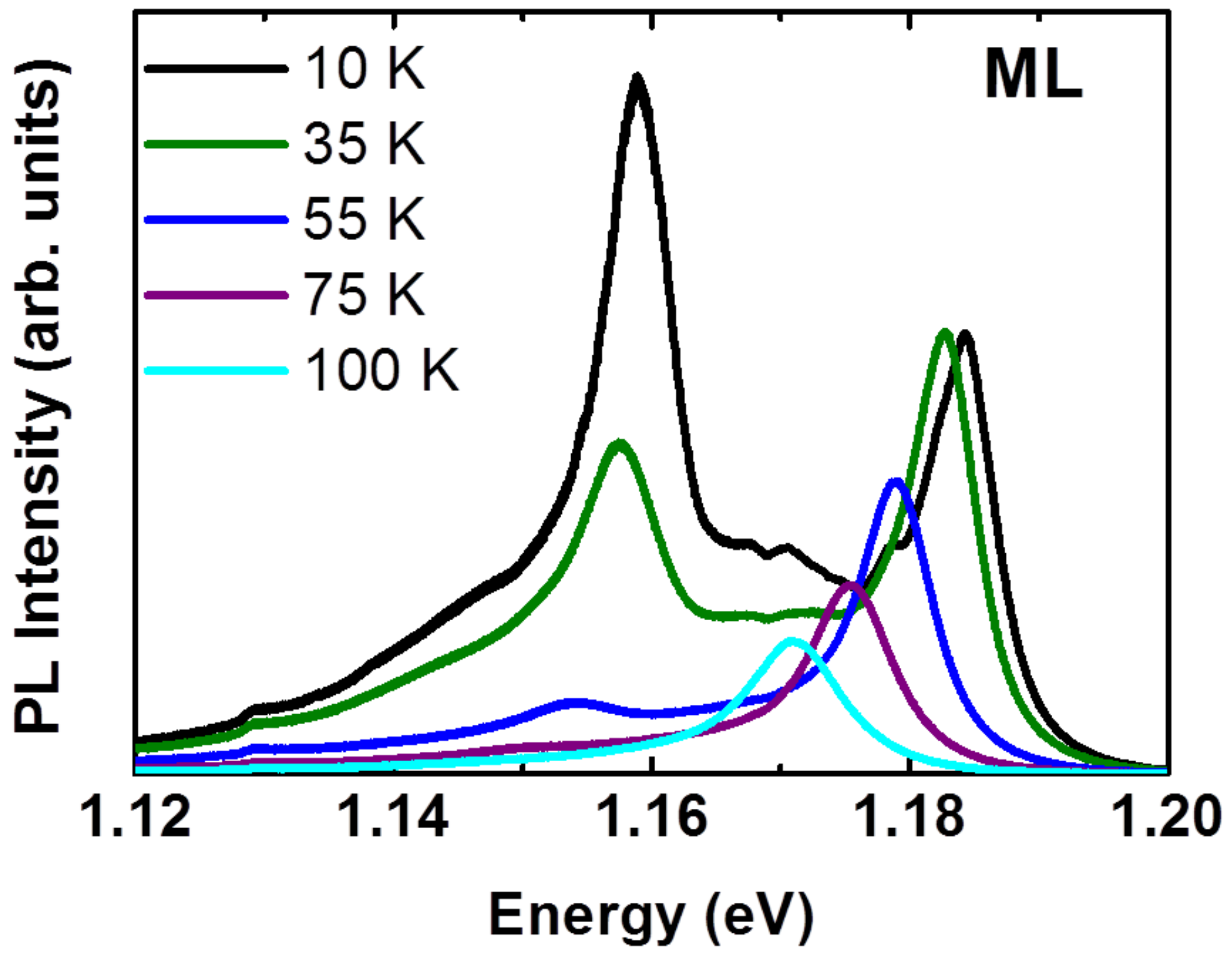

**FIG.3a:**

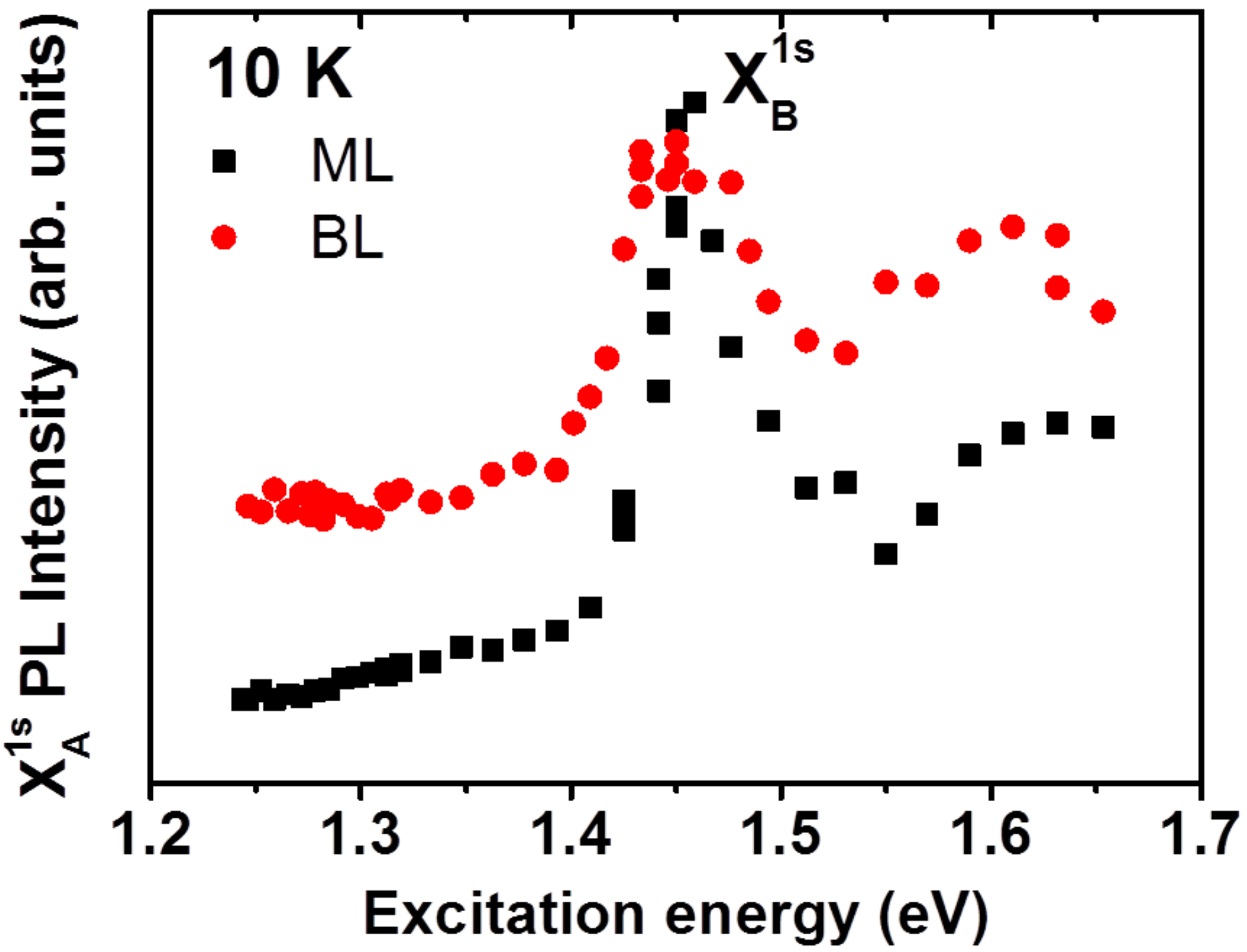

**FIG.3b:**

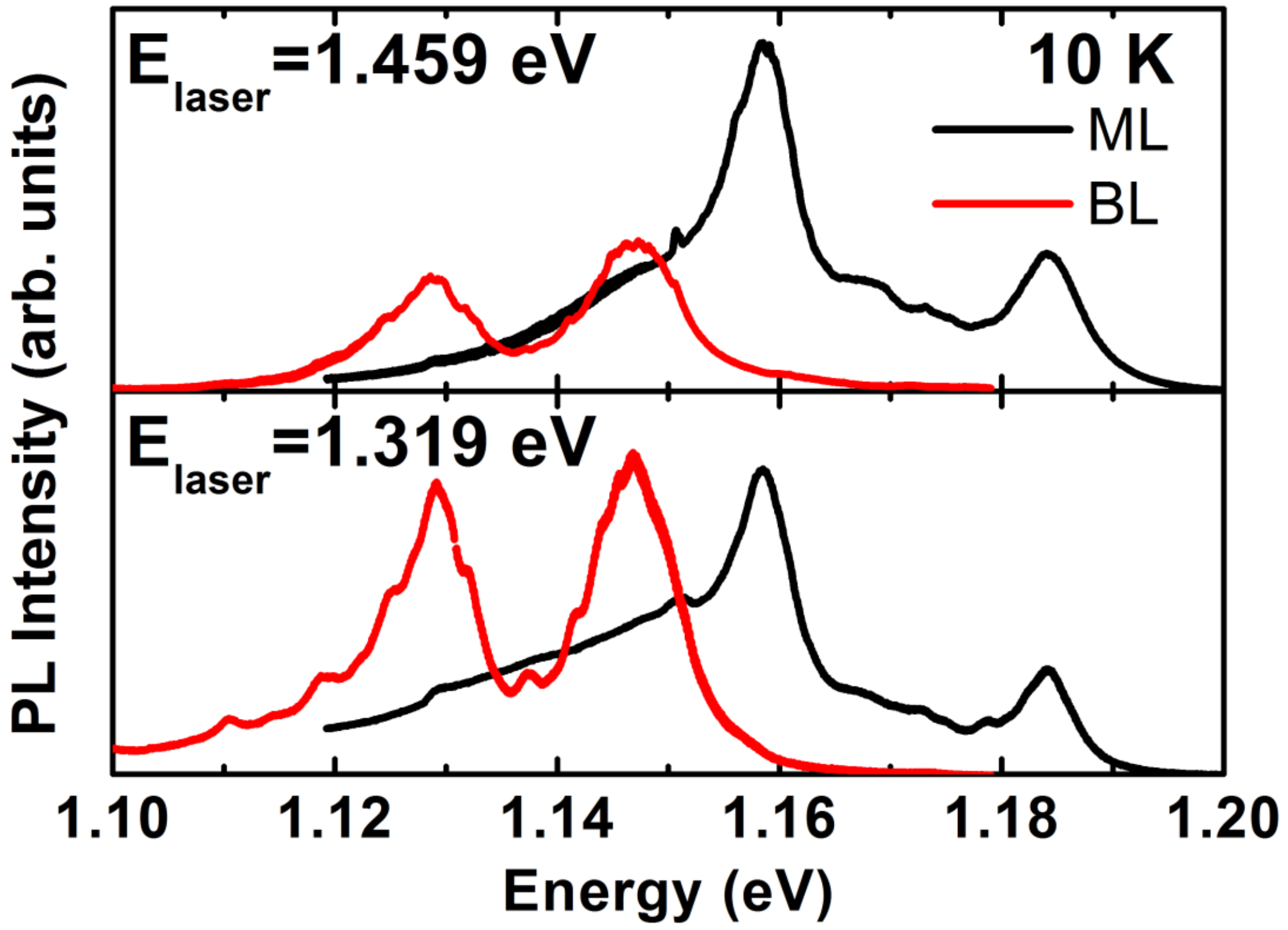

**FIG.4:**

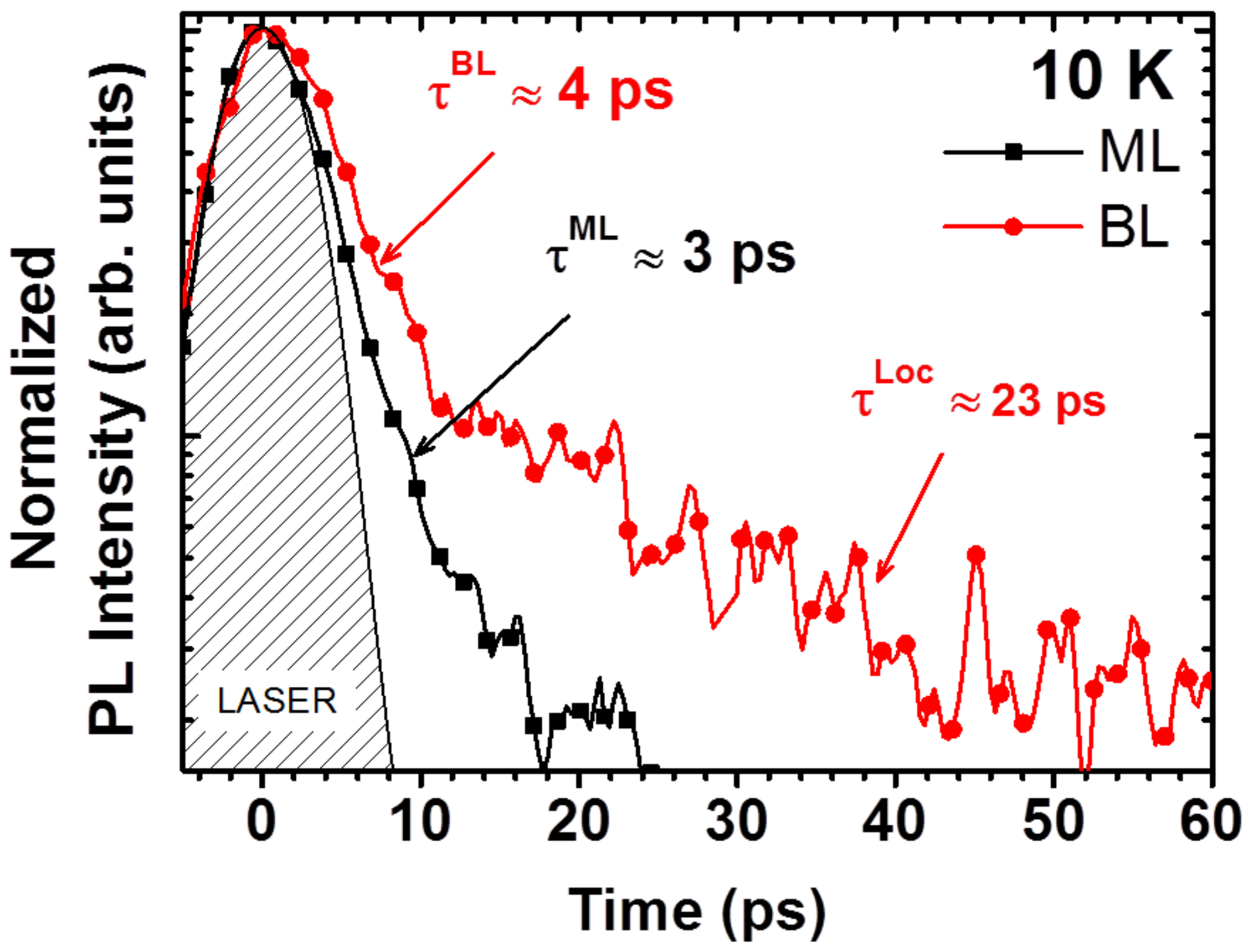

10 K
ML
BL
τBL ≈ 4 ps
τML ≈ 3 ps
τLoc ≈ 23 ps
LASER
Normalized
PL Intensity (arb. units)
0
10
20
30
40
50
60
Time (ps)

**FIG.5:**

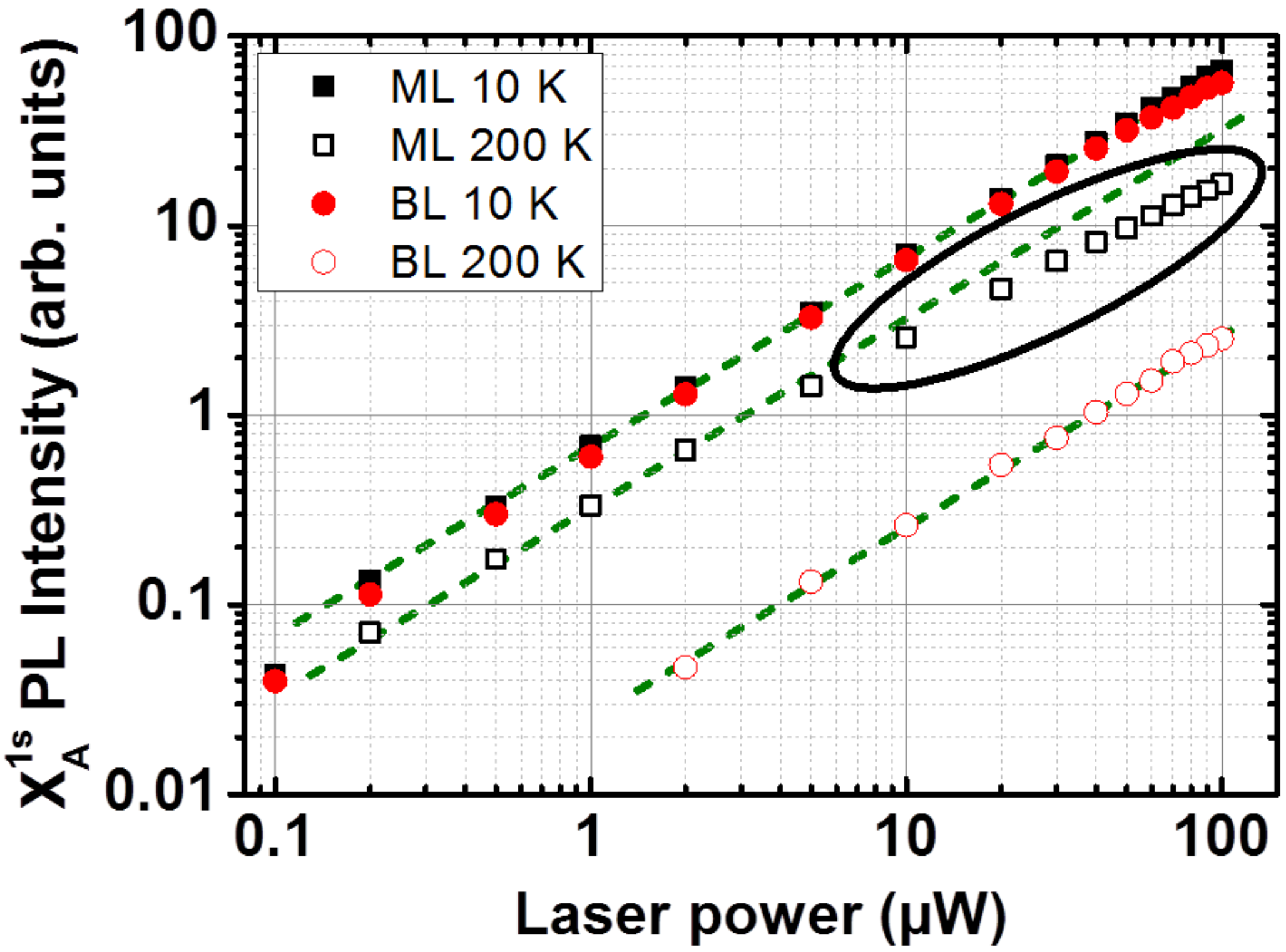

**Figure captions**

Fig. 1: Quasiparticle band structure for a freestanding $MoTe_2$ (a) ML and (b) BL, at the $G_0W_0$ level of theory, extracted from a Wannier localization procedure. The energy at the valence band maximum (in K) is set to zero for sake a comparison. Imaginary part of the transverse dielectric constant $\varepsilon_2^{xx}(\omega)$ as a function of photon energy (in eV) for $MoTe_2$ (c) ML and (d) BL. Red bars represent the relative oscillator strengths for the optical transitions, when the dashed line indicates the $G_0W_0$ band gap.

Fig. 2: (a) PL spectra of $MoTe_2$ ML (black) and BL (red) at 10 K. Inset: optical microscope image used for the identification of ML and BL regions. (b) PL spectra of $MoTe_2$ ML from 10 K to 100 K. The sample is excited with a cw HeNe laser (633 nm) at a power of 50 µW.

Fig. 3: (a) PLE spectra (PL intensity of the $X_A^{1s}$ peak as a function of excitation energy) of $MoTe_2$ ML (black square points) and BL (red circle points) at 10 K. The sample is excited with a cw Ti:Sa laser for wavelengths between 750 nm (1.65 eV) and 950 nm (1.31 eV) and with a tunable laser diode for wavelengths between 950 nm (1.31 eV) and 1000 nm (1.24 eV). The laser power is 50 µW. (b) PL spectra of $MoTe_2$ ML (black) and BL (red) at 10 K for two different excitation laser energies. The sample is excited at a power of 50 µW.

Fig. 4: Time-resolved photoluminescence of the neutral exciton $X_A^{1s}$ in $MoTe_2$ ML (black) and BL (red) following a resonant excitation on the B exciton (E[laser]= 1.458 eV (850 nm)) at 10 K. The instrument response is obtained by detecting the backscattered laser pulse on the sample surface, see the hatched area labeled “LASER”.

Fig. 5: Dependence of the $X_A^{1s}$ PL intensity with the laser power in $MoTe_2$ ML (black square points) and BL (red circles points) at 10 K (filled symbols) and 200 K (opened symbols). The sample is excited with a cw HeNe laser (633 nm). The green dashed lines indicate linear slopes. The black circle highlights the non-linear dependence of the $X_A^{1s}$ PL intensity in $MoTe_2$ ML at 200 K.